\documentclass{emulateapj}
\usepackage{apjfonts}

\newcommand{\HI}{\ion{H}{1}}

\slugcomment{The version accepted for publication in ApJ}
\shorttitle{Gamma-ray Emission from SNR W49B}
\shortauthors{Fermi LAT Collaboration}

\begin{document}

\title{Fermi-LAT Study of Gamma-ray Emission \\
in the Direction of Supernova Remnant W49B}

\author{
A.~A.~Abdo\altaffilmark{2,3}, 
M.~Ackermann\altaffilmark{4}, 
M.~Ajello\altaffilmark{4}, 
L.~Baldini\altaffilmark{5}, 
J.~Ballet\altaffilmark{6}, 
G.~Barbiellini\altaffilmark{7,8}, 
D.~Bastieri\altaffilmark{9,10}, 
K.~Bechtol\altaffilmark{4}, 
R.~Bellazzini\altaffilmark{5}, 
E.~D.~Bloom\altaffilmark{4}
% ...
}

% ...
\altaffiltext{2}{Space Science Division, Naval Research Laboratory, Washington, DC 20375, USA}
\altaffiltext{3}{National Research Council Research Associate, National Academy of Sciences, Washington, DC 20001, USA}
\altaffiltext{4}{W. W. Hansen Experimental Physics Laboratory, Kavli Institute for Particle Astrophysics and Cosmology, Department of Physics and SLAC National Accelerator Laboratory, Stanford University, Stanford, CA 94305, USA}
\altaffiltext{5}{Istituto Nazionale di Fisica Nucleare, Sezione di Pisa, I-56127 Pisa, Italy}
\altaffiltext{6}{Laboratoire AIM, CEA-IRFU/CNRS/Universit\'e Paris Diderot, Service d'Astrophysique, CEA Saclay, 91191 Gif sur Yvette, France}
\altaffiltext{7}{Istituto Nazionale di Fisica Nucleare, Sezione di Trieste, I-34127 Trieste, Italy}
\altaffiltext{8}{Dipartimento di Fisica, Universit\`a di Trieste, I-34127 Trieste, Italy}
\altaffiltext{9}{Istituto Nazionale di Fisica Nucleare, Sezione di Padova, I-35131 Padova, Italy}
\altaffiltext{10}{Dipartimento di Fisica ``G. Galilei", Universit\`a di Padova, I-35131 Padova, Italy}
% ...

\begin{abstract}
We present an analysis of 
the gamma-ray data obtained with 
the Large Area Telescope (LAT) onboard the \emph{Fermi Gamma-ray Space Telescope} 
in the direction of SNR W49B (G43.3$-$0.2). 
A bright unresolved gamma-ray source detected at a significance of $38\sigma$ 
is found to coincide with SNR W49B. 
The energy spectrum in the 0.2--200~GeV range gradually steepens toward high energies. The luminosity is estimated to be $1.5\times10^{36}$~($D/8$~kpc)$^2$~erg~s$^{-1}$ 
in this energy range. 
There is no indication that the gamma-ray emission comes from a pulsar. 
Assuming that the SNR shell is the site of gamma-ray production, 
the observed spectrum can be explained either by the decay of neutral $\pi$ mesons 
produced through the proton--proton collisions or by electron bremsstrahlung.
The calculated energy density of relativistic particles responsible for the LAT flux is estimated to be remarkably large,  
$U_{e,p}>10^4\ {\rm eV\ cm}^{-3}$, for either gamma-ray production mechanism.
\end{abstract}


\section{Introduction}

Galactic cosmic rays are widely believed to be accelerated in supernova remnants (SNRs) through the diffusive shock acceleration process \citep[e.g.,][]{BE87}. 
Several SNRs
have recently been detected with the Large Area Telescope (LAT) onboard the \emph{Fermi} Gamma-ray Space Telescope; specifically SNRs W51C, Cassiopeia~A, W44, and IC443 
 \citep{FermiW51C,FermiCasA,FermiW44,FermiIC443}. 
Except for Cas A, the LAT-detected SNRs 
 are known to be interacting with molecular clouds. 
 The GeV emission from such SNRs is expected to be dominated 
by  the hadronic gamma rays due to the decay of $\pi^0$ mesons, 
 since the ambient dense molecular cloud would enhance the proton-proton collisions  \citep{ADV94}. 
 The observed gamma-ray sources associated with cloud-interacting SNRs 
are all seen to be spatially extended in the LAT data.
Based on the extension and its comparison with radio data, it is concluded that  the gamma-ray  emission comes 
from SNRs not from pulsars/pulsar wind nebulae (PWNe).  
  The LAT spectra of these SNRs steepen above a few GeV. 
 Although electron bremsstrahlung cannot be ruled out,  $\pi^0$-decay  emission is the most plausible explanation for the observed LAT data \citep{FermiW51C,FermiW44}. 
The breaks in the observed spectra may be accounted for by an energy-dependent escape of accelerated protons at SNRs \citep{AA96}. 

The gamma-ray measurements in the TeV range provide direct support for the acceleration 
of particles up to $\sim 100$ TeV in SNR shells \citep{HESS1713n2}.
The TeV gamma rays in SNR RX~J1713.7$-$3946, one of the most prominent 
examples of TeV-emitting SNRs,  can be ascribed to 
the decay of $\pi^0$ mesons produced in {\it pp} collisions \citep[e.g.,][]{BV08} if 
the average magnetic field strength is larger than $\simeq 15\ \mu$G \citep{Uchi07}. 
However, the emission mechanism remains unsettled largely because of 
poorly constrained physical conditions in 
the gamma-ray-emitting zone in SNR RX~J1713.7$-$3946. 
Other examples  are valuable for discriminating the origins of the gamma-ray emission.

SNR W49B (G43.3$-$0.2) has a bright radio shell and centrally peaked thermal X-ray emission. The interaction between W49B and molecular clouds was evidenced by 
observations of  mid-infrared lines from shocked molecular hydrogen \citep{Spitzer}. 
\HI~Zeeman observations also suggest the interaction \citep{H1}. 
Near infrared [Fe II] emission exhibits filamentary structures, tracing radiative shocks
\citep{Keohane}. 
The age of W49B is estimated to be in the range of $\sim$1000--4000~yr \citep{Pye,Hwang}, and 
the  distance is estimated to be 8--11~kpc \citep{dist,Moffett,H1}.
The radio continuum map shows a shell structure with a diameter of $\sim 4\arcmin$ 
($\sim10$~pc at 8 kpc). 
The radio flux density is 38 Jy at 1~GHz. 
The radio emission is linearly polarized and the spectral index is $\alpha=0.48$~\citep{radioSNR} in the frequency range 0.3--30~GHz, indicating a synchrotron origin. 
No optical emission is detected from the source due to the severe extinction 
through the Galactic plane. 
Although the ATNF pulsar database~\citep{ATNF}\footnote{http://www.atnf.csiro.au/research/pulsar/psrcat} lists seven pulsars with the spin-down luminosity $>1\times 10^{34}$~erg s$^{-1}$ within 1\fdg 0 of the SNR position, no pulsar candidate has been reported within 0\fdg 4.
Prior to our LAT observations, gamma-ray emission had not been detected in the GeV or TeV bands.

Here we report the LAT observations in the direction of SNR W49B. 
A GeV gamma-ray source spatially coincident with W49B 
 is designated as 0FGL J1911.0+0905 in the initial source list published by the \emph{Fermi} LAT collaboration, which includes the 205 most significant sources based on the observation in the first three months \citep{BSL}. Is is also designated as 1FGL~J1910.9+0906c in the year-1 catalog (1FGL catalog)\footnote{available at the \emph{Fermi} Science Support Center. http://fermi.gsfc.nasa.gov/ssc}.
In this paper, we present a detailed analysis of this LAT source with much longer accumulation time of about 17 months.
This paper is organized as follows. In section 2, the observation and the data reduction are summarized. The analysis results for the LAT source in the direction of SNR W49B are reported using 17 months of the LAT data in section 3. In section 4, we discuss whether the gamma rays come from the SNR shell or a pulsar, and study the cosmic-ray acceleration using multi-wavelength data.

\section{Observation and Data Reduction}
The \emph{Fermi Gamma-ray Space Telescope} was launched on June 11 2008. The Large Area Telescope (LAT) onboard \emph{Fermi} is composed of electron-positron pair trackers, featuring solid state silicon trackers and cesium iodide calorimeters, sensitive to photons in a very broad energy band (from 0.02 to $>$300~GeV). The LAT has a large effective area ($\sim$8000 cm$^2$ above 1 GeV if on-axis), viewing $\sim$2.4 sr of the full sky with a good angular resolution (68 \% containment radius better than $\sim$1$^{\circ}$ above 1~GeV). 
The tracker of the LAT is divided into \emph{front} and \emph{back} sections.
The front section (first 12 planes) has thin converters to improve the PSF, while the back section (4 planes after the front section) has thicker converters to enlarge the effective area.
The angular resolution of the back events is a factor of two worse than that of the front events at 1~GeV. 

The LAT data used here were collected for about 17 months from August 4 2008 to December 26 2009. The 
\emph{diffuse} event class was chosen and photons beyond the earth zenith angle of 105$^\circ$ were excluded to minimize Earth albedo gamma rays.

Among the standard science analysis tools\footnotemark[2],
we utilized {\sf gtlike} for spectral fits and {\sf gtfindsrc} to find a point source location. 
With {\sf gtlike}, an unbinned maximum likelihood fit is performed 
on the spatial and spectral distributions of observed gamma rays to optimize spectral parameters of the input model taking into account the energy dependence of the point spread function (PSF).
On the other hand, 
{\sf gtfindsrc} optimizes a point source location by finding the best likelihood for different positions around an initial guess until the convergence tolerance for a positional fit is reached.
The {\sf P6\_V3} instrument response functions were used for the analyses in this paper.
Details of the LAT instrument and data reduction are described in \citet{LAT}.

\section{Analysis and Results}

\subsection{Detection and Source Localization}\label{sed:detect}

\begin{figure}[htbp] 
 \epsscale{1.0}
\plotone{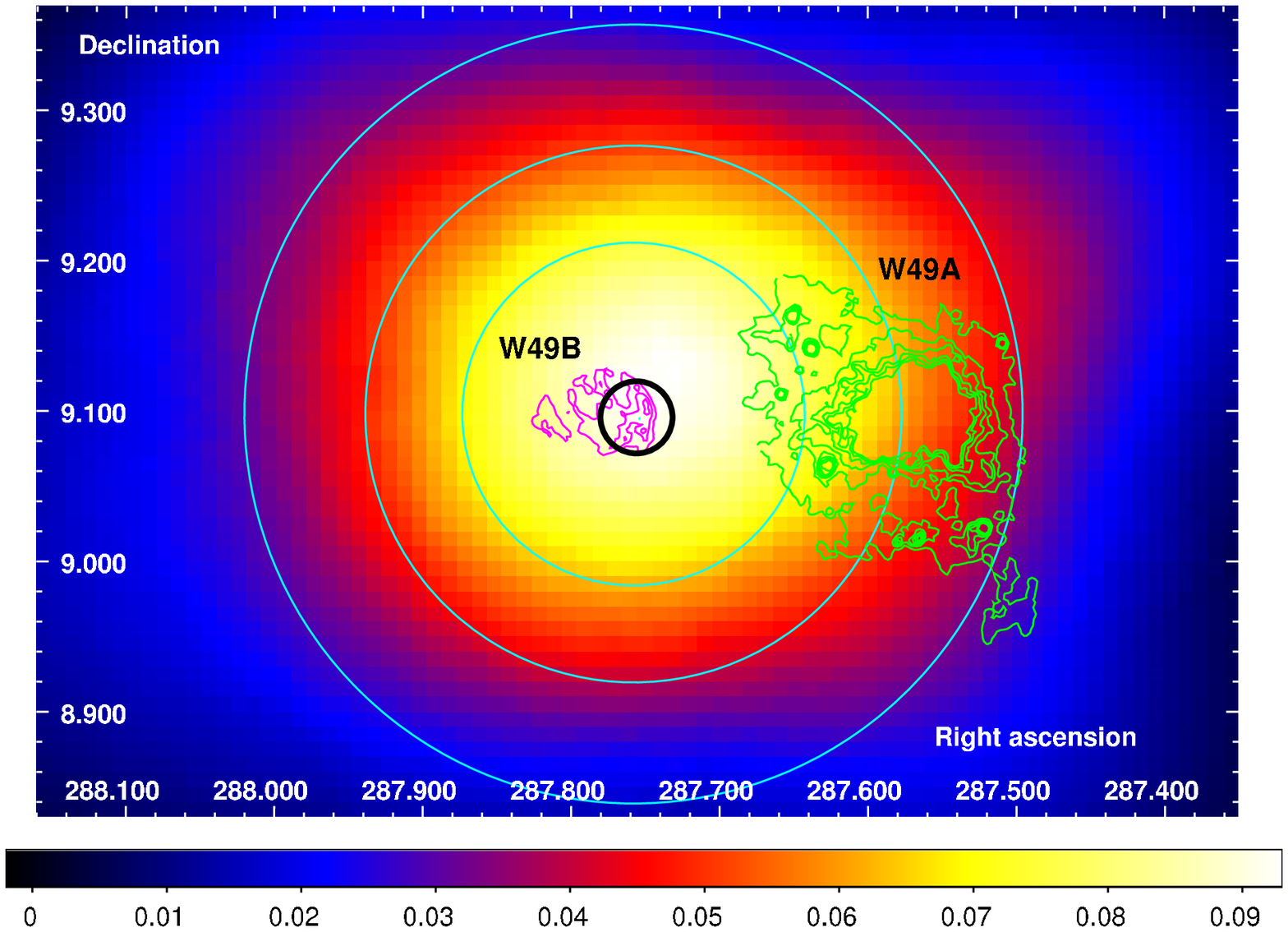}
\plotone{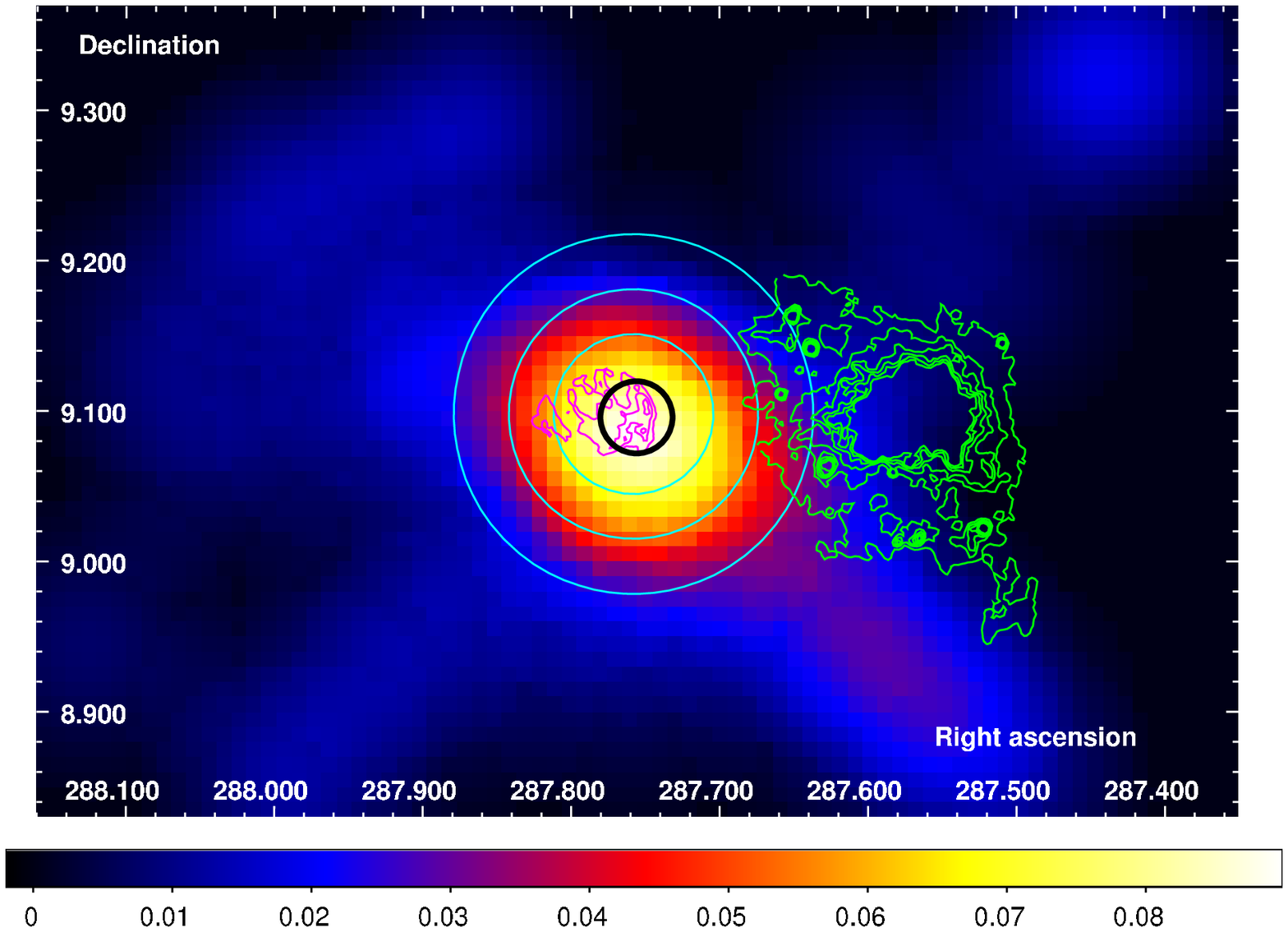}
\caption{\small 
\emph{Fermi} LAT count map in the vicinity of SNR W49B in units of counts per pixel. 
The pixel size is $0\fdg 01$. 
The LAT localization is represented by a black circle with a radius of $0\fdg 024$ (95 \% confidence level) centered at ($\alpha$, $\delta$)=($287\fdg 756$, $9\fdg 096$). 
Cyan circles represent radii of the effective LAT PSF at 75\%, 50\% and 25\% of the peak.
Magenta and green contours indicate W49B and W49A in the Spitzer IRAC 5.8 $\mu$m, respectively.
(Top) The count map in 2--6 GeV is smoothed by a Gaussian kernel of $\sigma =0\fdg 2$.
(Bottom) The count map in 6--30 GeV is smoothed by a Gaussian kernel of $\sigma =0\fdg 1$.
\label{smallFOV}}
\end{figure}

The LAT observation revealed significant ($38\sigma$) gamma-ray emission from the direction of SNR W49B with 17 months of data.
Figure~\ref{smallFOV} shows LAT count maps in the vicinity of SNR W49B in the 2--6~GeV and 6--30~GeV bands. 
Only \emph{front} events are used in the count map to achieve better angular resolution.
The effective LAT point-spread function (PSF) is constructed using a spectral shape obtained through a maximum likelihood fit ({\sf gtlike)} in the corresponding energy band for each count map (see \S\ref{sec:SED}).
The statistical and systematic uncertainties in the spectral shape do not noticeably affect the PSF shape.
A \emph{Spitzer} near-infrared (5.8~$\mu$m) map, which traces ionic shocks in the SNR, 
is overlaid on the count maps\footnote{The IR data are available from NASA/IPAC Infrared Science Archive.\\
http://irsa.ipac.caltech.edu/data/SPITZER/GLIMPSE}.
Both count maps clearly suggest that gamma-ray emission comes predominantly from the SNR W49B region, not from a nearby star forming region, W49A.
Comparisons between gamma-ray distributions and LAT PSFs in both energy bands 
indicate that the observed gamma-ray emission could be consistent with a point source. 

\begin{figure}[htbp] 
\epsscale{1.0}
\plotone{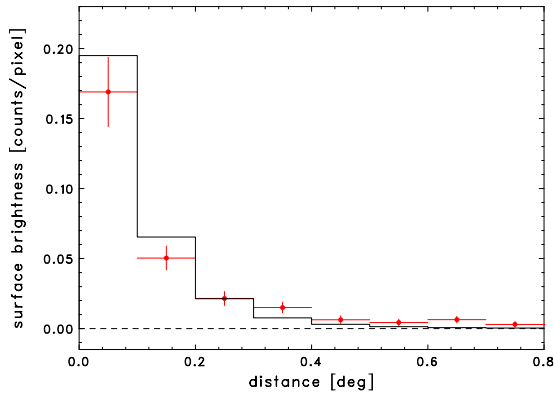}
\caption{\small
The radial profile of the LAT count map with \emph{front} data in 2 to 30 GeV in units of 
counts per pixel. 
The pixel size is $0\fdg 01$.
The origin of the profile is the LAT localization of the W49B source at ($\alpha$, $\delta$)=($287\fdg 756$, $9\fdg 096$).
The histogram shows the profile of the effective LAT PSF.
\label{radprof}}
\end{figure}

In order to confirm the consistency with a point source, a radial profile of the gamma rays from the above source location is compared with that expected for a point source for \emph{front} events in 2--30~GeV band as shown in Figure~\ref{radprof}.
The background, which is composed mainly of the Galactic diffuse emission, is subtracted. 
No sign of spatial extension can be seen in Figure~\ref{radprof}.

To evaluate the consistency with a point source quantitatively, we compared the likelihood of the spectral fit for a point source and an elliptical shape ($3'\times 4'$ in size; compatible with the extent of the IR image as shown in Figure~\ref{smallFOV}) with a uniform surface brightness. 
Here, we assumed a broken power law function to model the source spectrum in the fit (see \S\ref{sec:SED} for details).
The resulting likelihood was almost the same for both cases (difference of log likelihood was $\sim 3$), which means the source emission is consistent with that from a point source.
Therefore, to simplify the analyses, the gamma-ray source in the SNR W49B region is analyzed as a point source in this paper. 
Assuming a point source, the gamma-ray source position was found to be ($\alpha$, $\delta$)=($287\fdg 756$, $9\fdg 096$) with an error radius of $0\fdg 024$ at 95\% confidence level using {\sf gtfindsrc}, as indicated by the black circle in Figure~\ref{smallFOV}.

%%%%%%%%%%%%%%%%%%%%%%%%%%%%%%%%%%%%
%%%%%%%%%%%%%%%%%%%%%%%%%%%%%%%%%%%%

%%%%%%%%%%%%%%%%%%%%%%%%%%%%%%%%%%%%
%%%%%%%%%%%%%%%%%%%%%%%%%%%%%%%%%%%%

\subsection{Evaluation of Galactic Diffuse Model}\label{sec:galdif}

\begin{figure}[htbp] 
\epsscale{1.0}
\plotone{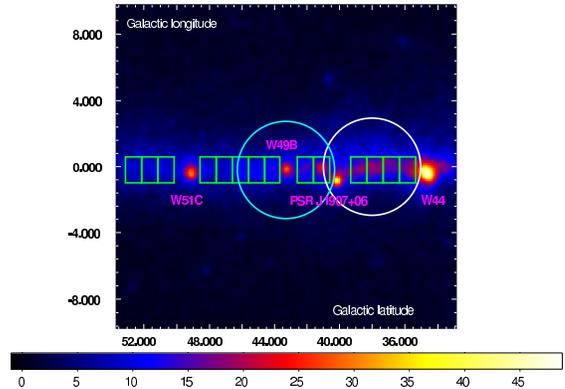}
\caption{\small
The LAT count map above 1~GeV around SNR W49B in units of counts per pixel. 
The pixel size is $0\fdg 1$, and Gaussian smoothing is applied with a kernel size of $\sigma = 0\fdg 3$. The W49B source, which is located at the center, is clearly visible. 
The green boxes ($1\fdg0 \times 1\fdg6$) represent the regions used for the evaluation of spatial dispersion of the difference between the Galactic diffuse model and the observed distribution. 
The cyan and white circles represent the regions where the flux of the Galactic diffuse model was varied to evaluate effects of spatial dispersion of the model.
\label{segments}}
\end{figure}

Since uncertainties associated with the underlying Galactic diffuse emission are expected to be the largest systematic effects for spectral analyses of the W49B source, those effects should be carefully evaluated. 
The uncertainties of the Galactic diffuse emission are primarily due to the imperfection of the Galactic diffuse model and/or the contributions from unresolved point sources.
As a first step of the evaluation process, the position and energy dependences of the discrepancies between the observed gamma-ray distributions and the Galactic diffuse model are studied in the regions where the Galactic diffuse emission is considered to be dominant around the W49B source.
The normalization of the Galactic diffuse model is determined by 
running {\sf gtlike} for a circular  region with a radius of $10^\circ$ centered on the W49B source  in the energy range of 0.2--200~GeV.
The position of the W49B source is fixed at ($\alpha$, $\delta$)=($287\fdg 756$, $9\fdg 096$) determined by {\sf gtfindsrc} (see \S~\ref{sed:detect}).
The positions and spectral shapes of all other sources are fixed at the value in the 1FGL catalog,
while the flux is allowed to vary, except for PSR J1907+06 \citep{PSRJ1907}, 
SNR W51C \citep{FermiW51C} and SNR W44 \citep{FermiW44} which are 3, 6 and 9 degrees away from W49B, respectively. 
Since these sources around SNR W49B are very bright as evident in Figure~\ref{segments}, we carefully evaluated spectral models for these sources.
For this study we modeled W44 as two point sources at ($\alpha$, $\delta$)=($283\fdg 89$, $1\fdg 56$) and ($284\fdg 10$, $1\fdg 15$), to approximately account for its angular extent, 
while the positions of the other sources are fixed at the values determined by the catalog. 
The spectral shape of these four bright sources is assumed to be a broken power law since likelihood tests between a power law function and a broken power law function favored a broken power law hypothesis at $> 10\sigma$ (PSR J1907+06), $> 7\sigma$ (SNR W51C), 
$> 15\sigma$ (SNR W44) confidence levels.
All spectral parameters (flux, spectral break and spectral indices at low and high energy) are allowed to vary in the fitting since the spectral model is different from that reported in the 1FGL catalog.
The Galactic diffuse emission is modeled using ``gll\_iem\_v02.fit".
An isotropic component (isotropic\_iem\_v02.txt) is also included to account for instrumental and extragalactic diffuse backgrounds. 
The both background models are the standard diffuse emission models released by the LAT team\footnotemark[2].
The normalization factors of the Galactic diffuse  and the isotropic models are allowed to vary. 

%%%%%%%%%%%%%%%%%%%%%%%%%%%%%%%%%%%%
%%%%%%%%%%%%%%%%%%%%%%%%%%%%%%%%%%%%

\begin{figure}[htbp] 
\epsscale{1.0}
\plotone{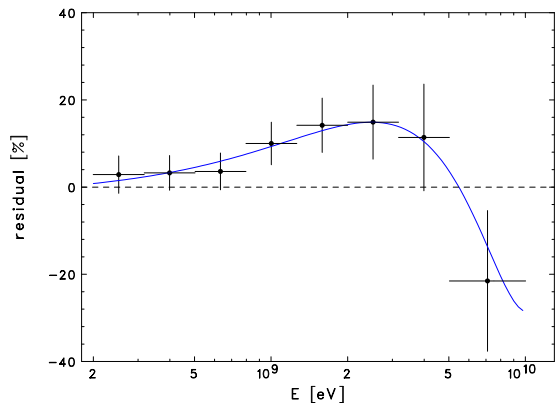}
\caption{\small
The fractional residuals at 8 energy bins in 0.2 to 10 GeV between the observed LAT data and 
the best-fit Galactic diffuse emission model in the nearby circle centered at ($\alpha$, $\delta$)=($288\fdg 4$, $10\fdg 2$) with radius of $0\fdg 5$.
The fluxes of all sources included in the fit-model except for the Galactic diffuse component are subtracted from the observed data. 
The blue line shows a cubic function fitted to the residual data.
\label{deltaE}}
\end{figure}

In order to evaluate the the validity of spectra for background models, we compared two counts spectra in the 0.2--10~GeV band;
the spectrum expected from the models obtained by the above procedure, and the observed spectrum. 
This comparison is performed in a nearby circular region with a radius of $0\fdg 5$ centered on $\Delta l \sim+2^\circ$ and $\Delta b \sim0^\circ$ from SNR W49B where the Galactic diffuse component is dominant. 
Figure~\ref{deltaE} shows resulting fractional residuals, namely (observed$-$model)/model, as a function of energy. 
We fit the residuals with a cubic function as shown in Figure~\ref{deltaE}, which will be used to estimate the systematic error in flux due to uncertainties of Galactic diffuse model as discussed in \S~\ref{sec:SED}.

%%%%%%%%%%%%%%%%%%%%%%%%%%%%%%%%%%%%
%%%%%%%%%%%%%%%%%%%%%%%%%%%%%%%%%%%%

\begin{figure}[htbp] 
\epsscale{1.0}
\plotone{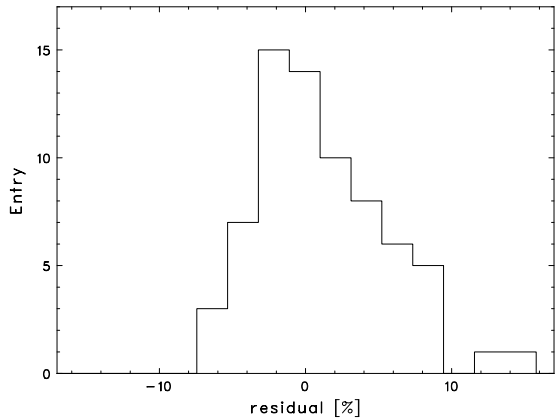}
\caption{\small
The histogram of the fractional residuals between the observed LAT data and the Galactic diffuse model fixed at the best-fit parameters determined by {\sf gtlike}. 
The fluxes of all sources included in the fit-model except for the Galactic diffuse component are subtracted from the observed data.
The residual was calculated in 5 energy bands (0.20--0.32 GeV, 0.32--0.50 GeV, 0.50--0.80 GeV, 0.80--1.3, 1.3--10 GeV) for each region as shown in Figure~\ref{segments} .
\label{deltaSp}}
\end{figure}

Uncertainties of the spatial distribution of the Galactic diffuse emission are evaluated 
by measuring the dispersion of the fractional residuals in 14 regions, where  the Galactic diffuse component is dominant (Figure~\ref{segments}).
The regions around 4 very bright sources, the W49B source, PSR J1907+06, SNR W51C and SNR W44, are excluded.
The fractional residual for each region is calculated in 5 energy bands,
0.20--0.32 GeV, 0.32--0.50 GeV, 0.50--0.80 GeV, 0.80--1.3, 1.3--10 GeV.
Figure~\ref{deltaSp} shows the resulting distribution of the fractional residuals for 14 regions in 5 energy bands.
The figure shows that 68\% and 90\% of the fractional residual are within 4\% and 6\%, respectively. 
To be conservative, the fractional residual of 6\% will be used to estimate the systematic error in flux due to uncertainties of Galactic diffuse model below.

%%%%%%%%%%%%%%%%%%%%%%%%%%%%%%%%%%%%
%%%%%%%%%%%%%%%%%%%%%%%%%%%%%%%%%%%%

\subsection{Gamma-ray Spectrum of W49B}\label{sec:SED}

\begin{figure}[htbp] 
\epsscale{1.0}
\plotone{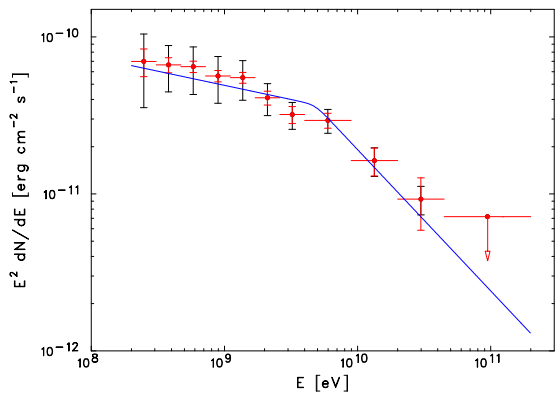}
\caption{\small
Spectral energy distribution of the W49B source measured with the \emph{Fermi} LAT. 
The vertical red and black lines represent statistical errors (1$\sigma$) and systematic errors, respectively.
The blue line represents the best-fit broken power law from an unbinned likelihood fit in 0.2--200 GeV.
\label{SED}}
\end{figure}

The spectral energy distribution (SED) of the source associated with W49B is evaluated by dividing the 0.2--200~GeV energy band into 11 energy bins, extracting data inside a circular region with a radius of $10^\circ$ centered on the W49B source and by using {\sf gtlike} to obtain a flux value at the center of each bin.
In each {\sf gtlike} run, the W49B source, the other 1FGL sources, Galactic diffuse and isotropic backgrounds 
are fitted with their normalization free. The W49B source is fitted with a simple power--law function in each energy bin with its spectral index fixed at 2.2 below 5~GeV and 2.9 above 5~GeV using the fitting result in 0.2--200~GeV (see below), while the indices of the other sources are fixed at the values in the 1 FGL catalog. Note that the obtained flux of the W49B source is insensitive to the choice of the index, if it is fixed in a reasonable range (say, 2--3).
Figure~\ref{SED} shows the resulting SED for the W49B source. 

%%%%%%%%%%%%%%%%%%%%%%%%%%%%%%%%%%%%
%%%%%%%%%%%%%%%%%%%%%%%%%%%%%%%%%%%%

In order to evaluate systematic effects on the SED due to uncertainties of the Galactic diffuse model, we varied the Galactic diffuse model used in the fit. 
Systematic errors due to uncertainties in the energy spectrum of the Galactic diffuse model are estimated by comparing the fit 
with and without the modification of the energy distributions of the Galactic diffuse model according to the curve in Figure~\ref{deltaE}.
We did not modify the shape of the energy distribution above 10~GeV.
Though the fractional residual intensities in Figure~\ref{deltaE} are within our current understanding of the systematic uncertainties in the effective area, these residuals were used conservatively as uncertainties of the Galactic diffuse model.
We obtain an estimate of uncertainties as:
 $\leq$ 30\% for below 1~GeV, $\leq$ 20\% in 1--2~GeV, and $\leq$10\% above 3~GeV.
Systematic errors due to uncertainties of spatial distribution of the Galactic diffuse model as shown in Figure~\ref{deltaSp} are estimated, 
using two modified Galactic diffuse models in which fluxes are varied by 6\% in all energy bins for one of two regions with a 3 degree radius, 
a disk centered on W49B or an offset disk at ($\alpha$, $\delta$)=($285\fdg 18$, $4\fdg 51$). 
The resulting systematic errors are estimated to be 45\% at 300~MeV, decreasing to 12\% at 700~MeV, and $\leq$6\% above 1~GeV for both cases.
We adopt the maximum value among these errors at each energy bin as the systematic error due to the Galactic diffuse model.
Other systematic errors include uncertainties of the effective area which 
are 10\% at 100 MeV, decreasing to 5\% at 560 MeV, and increasing to 20\% at 10 GeV and above.
Total systematic errors are set by adding in quadrature the uncertainties due to the Galactic diffuse model and the effective area.
The total systematic errors in each energy bin are indicated by black error bars in Figure~\ref{SED} while statistical errors are indicated by red error bars.

\begin{figure}[htbp] 
\epsscale{1.0}
\plotone{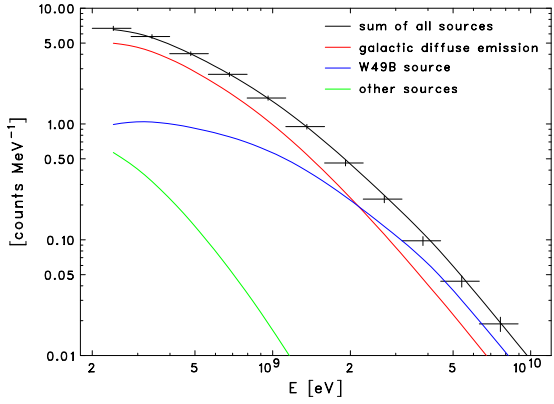}
\caption{\small
The count spectrum within a radius of $0\fdg 5$ around the W49B source location at ($\alpha$, $\delta$)=($287\fdg 756$, $9\fdg 096$). 
The blue, red, green and black line shows best-fit model curves for the W49B source, the Galactic diffuse emission model, the sum of the other sources (all sources except for the W49B source and the Galactic diffuse model) and the sum of all sources included in the fit-model, respectively. 
\label{spec_cnt}}
\end{figure}

Inspection of Figure~\ref{SED} suggests a spectrum steepening above a few GeV. We performed a likelihood-ratio test between a power-law (the null hypothesis) and 
a smoothly broken power-law functions (the alternative hypothesis) for 0.2--200~GeV data inside a circular  region with a radius of $10^\circ$ centered on the W49B source.
The smoothly broken power-law function is described as
\begin{equation}
\frac{dN}{dE}=KE^{-\Gamma_1}\left(1+\left(\frac{E}{E_{\rm break}} \right)^{\frac{\Gamma_2-\Gamma_1}{\beta}} \right)^{-\beta}, 
\label{eq:cutoff}
\end{equation}
where photon indices $\Gamma_1$ below the break, $\Gamma_2$ above the break, a break energy ${E_{\rm break}}$ 
and a normalization factor $K$ 
are free parameters. The parameter $\beta$ is fixed at 0.05. The simple broken power-law function is not adopted here, 
since the function cannot be differentiated at the break energy resulting in unstable fit results and inaccurate error estimates.
We obtained a test statistics of ${\rm TS_{BPL}} = -2\ln (L_{\rm PL}/L_{\rm BPL}) = 22.9$, which means a simple power law can be rejected at a significance of 4.4$\sigma$. The parameters obtained with the broken power law model are photon indices $\Gamma_1=2.18 \pm 0.04$, $\Gamma_2=2.9\pm0.2$, and $E_{\rm break}=4.8\pm1.6$ GeV, with 
an integrated flux in 0.2--200 GeV of $(1.74\pm0.06)\times10^{-7}$~photon~cm$^{-2}$~s$^{-1}$, while the photon index obtained with the simple power law is $2.29\pm0.02$.
The gamma-ray luminosity in 0.2--200 GeV is calculated as 
  $1.5\times10^{36}~(D/8$~kpc)$^2$~erg~s$^{-1}$.
Figure~\ref{spec_cnt} shows the resulting fit with a broken power-law spectrum to the count spectrum within a radius of $0\fdg 5$ around the W49B source location. 
This underscores the importance of understanding the Galactic diffuse emission for the spectral analyses of the W49B source.
We checked if  the significance of the spectral break changes for different Galactic diffuse models. 
We found that ${\rm TS_{BPL}}$ is 20.0 with the Galactic diffuse models used for evaluating the spatial distribution uncertainties, corresponding to a significance of 4.1$\sigma$. 
${\rm TS_{BPL}}$ is 11.8 for the Galactic diffuse model used for evaluating uncertainties of the energy spectrum, corresponding to a significance of 3.0$\sigma$.
Depending on the chosen Galactic diffuse model, the significance of the break ranges between 3 and 4.4$\sigma$.

%%%%%%%%%%%%%%%%%%%%%%%%%%%%%%%%%%%%
%%%%%%%%%%%%%%%%%%%%%%%%%%%%%%%%%%%%

\subsection{Upper limit on W49A}\label{sec:W49A}

W49A (G43.0+0.0) is one of the most active and luminous star forming regions ($\sim10^7~L_{\odot}$) in the Galaxy~\citep{Conti}, located $0\fdg 21$ to the west of SNR W49B as shown in Figure~\ref{smallFOV}. Its distance is estimated to be $11.4\pm1.2$~kpc~\citep{Gwinn}. 

In this analysis, we find no gamma-ray counterpart for W49A. An upper limit 
to the GeV flux from W49A is determined by performing {\sf gtlike} analysis. 
The model used for the fit includes W49A, the W49B source, all other 1FGL sources, Galactic diffuse and isotropic backgrounds. The W49A source is assumed to have uniform surface brightness inside a circle with radius $5'$. 
A simple power-law function with its photon index fixed at 2.0 or 2.5 is used to model the W49A spectrum. 
The upper limits on the flux (0.2--200~GeV) obtained from the fits with the indices fixed at 2.0 and 2.5 are $9.5\times 10^{-9}$~photon~cm$^{-2}$~s$^{-1}$ and $3.4\times 10^{-8}$~photon~cm$^{-2}$~s$^{-1}$ at 95\% confidence level, 
corresponding to luminosity limits of $<3\times 10^{35}\ (D/11.4\ {\rm kpc})^2\ {\rm erg\ s}^{-1}$ and $<4.9\times 10^{35}\ (D/11.4\ {\rm kpc})^2\ {\rm erg\ s}^{-1}$, respectively.
The uncertainties due to the Galactic diffuse model as discussed in \S\ref{sec:galdif} have little effect on the upper limit in the case of the photon index 2.0, while those increase the upper limit to $<7.8\times 10^{35}\ (D/11.4\ {\rm kpc})^2\ {\rm erg\ s}^{-1}$ in the case of the photon index 2.5.

\section{Discussion}\label{sec:discuss}

\subsection{Pulsar?}\label{sec:origin}

The gamma ray emission positionally coincident with SNR W49B 
is unresolved with the LAT. This is reasonable given the fact that 
the angular extent of SNR W49B is somewhat smaller than the effective LAT PSF.
The extent of GeV gamma-ray emission from middle-aged SNRs W51C \citep{FermiW51C} and
W44 \citep{FermiW44} made it possible to attribute 
the observed gamma-ray signals to the shells of these SNRs.
Since this is not possible with W49B, we will examine a possibility that a pulsar's magnetosphere is responsible for the observed gamma-ray emission even though no pulsed emission has been detected with the LAT. In addition, no radio pulsars are found within a radius of 0\fdg 4 around the LAT position of the W49B source in the ATNF catalog, while the LAT position is determined with 0\fdg 024 at 95\% confidence level. Note that we do not consider a PWN here, since the observed gamma-ray flux is very difficult to be accounted for by a radio-quiet PWN.

To compare the spectral shape of the W49B source with that of typical LAT pulsars in the first pulsar catalog~\citep{FermiP-ctlg},  
we  fit the LAT spectrum of the W49B source by a power law with an exponential cutoff:
%%%%%%%%%%%%
% a dummy line  %%%
%%%%%%%%%%%%
\begin{equation}
\frac{dN}{dE}=KE^{-\Gamma}\exp \left(-\frac{E}{E_{\rm cutoff}} \right), 
\label{eq:cutoff}
\end{equation}
where 
photon index $\Gamma$, a cutoff energy ${E_{\rm cutoff}}$ and a normalization factor $K$ 
are free parameters. 
The parameters of the W49B source obtained by {\sf gtlike} are $\Gamma=2.10 \pm 0.02$ and $E_{\rm cutoff} = 15 \pm 1$ GeV. 
We performed a likelihood-ratio test between a power law (the null hypothesis) and a cutoff power law (the alternative hypothesis) and obtained test statistics of ${\rm TS_{cutoff}} = -2\ln (L_{\rm PL}/L_{\rm cutoff}) = 27$, which means that we can reject a simple power law at a significance
of $\sim 5\sigma$. 
About 90\% of the 46 LAT pulsars in the catalog~\citep{FermiP-ctlg} have $\Gamma < 1.9$ and 
$E_{\rm cutoff} < 5.0$ GeV. 
No pulsars exhibits  $E_{\rm cutoff} > 6.5$ GeV 
among the LAT pulsars that have an error on $E_{\rm cutoff}$ less than 4~GeV.
The LAT spectrum of W49B is different from what has been obtained for almost all gamma-ray pulsars so far.

A pulsar may have eluded detection in X-rays due to the presence of bright X-ray 
emission from shock-heated plasmas. 
Using 55 ks of \emph{Chandra} data (PI: Holt, S. S.) 
we put an upper limit on the X-ray flux of a possible hidden pulsar 
of $F_{X}$ (2--10 keV) $< 6.5\times10^{-14}$~erg~s$^{-1}$~cm$^{-2}$ 
on the assumption that the pulsar spectrum is a power law with a photon index of 2.0. 
The foreground column density N$_{\rm H}$ used here is $6 \times 10^{22}$~cm$^{-2}$.
This corresponds to an upper limit on 
the X-ray  luminosity of $L_{X}$ (2--10 keV) $< 5\times10^{32}~(D/8$~kpc)$^2$ erg s$^{-1}$.
The empirical correlation of the X-ray and spin-down luminosity of rotation powered pulsars can be written as: 
 %%%%%%%%%%%%
% a dummy line  %%%

%%%%%%%%%%%%
\begin{equation}
\log L_{\rm X} = 1.34 \log L_{\rm sd} - 15.34, 
\label{eq:relation}
\end{equation}
where $L_{\rm X}$ and $L_{\rm sd}$ are the X-ray luminosity in the 2--10 keV and the spin-down energy loss in units of erg s$^{-1}$, respectively \citep{Possenti}. 
This relation constrains the spin down luminosity of any undetected pulsars in W49B to be 
$L_{\rm sd} < 1\times 10^{36}~(D/8$~kpc)$^2$\ ${\rm erg\ s^{-1}}$. 
However, 
the gamma-ray luminosity (0.2--200~GeV) of the W49B source is
$1.5 \times 10^{36}~(D/8$~kpc)$^2$\ ${\rm  erg\ s^{-1}}$, which exceeds $L_{\rm sd}$. 
Together with the spectral argument, 
we conclude that the gamma-ray emission in the direction of W49B is unlikely to 
come from a pulsar.

\subsection{Gamma Rays from the SNR Shell}\label{sec:mech}

\begin{figure}[htbp] 
\epsscale{1.0}
\plotone{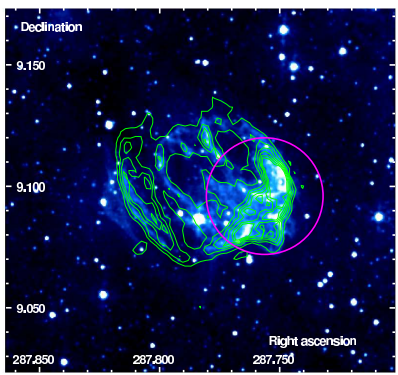}
\caption{\small  
LAT source position at a 95~\% confidence level (a magenta circle) 
is superposed on the \emph{Spitzer} IRAC 5.8~$\mu$m image.
Contours show 20~cm radio intensity obtained from MAGPIS~\citep{20cm}.
\label{fig:multi-img}}
\end{figure}

Here we consider a scenario in which  the gamma-ray source originates in 
the radio-emitting  shell of  SNR W49B.
This scenario is supported by 
the best-fit LAT position being coincident with the brightest part  of synchrotron radio emission as shown in Figure~\ref{fig:multi-img}. 
The near-infrared [Fe II] emission, arising from warm ionized gas with a density of 
order $1000\ {\rm cm}^{-3}$, correlates well with the synchrotron map \citep{Keohane}.

We assume that the particles responsible for the LAT flux are distributed 
in a radio-emitting zone which can be characterized 
by a constant hydrogen density $n_{\rm H}$ and 
magnetic field strength $B$. 
The volume of the emission zone is written as $V = f (4\pi/3) R^3$ where 
$f \leq 1$ denotes a filling factor and $R= 4.4\ \rm pc$ is the radius of the remnant. 
The radio-emitting material would originate in swept-up stellar wind and/or 
interstellar gas. 
We adopt the total mass contained in the zone as $M_{\rm H} = 50M_\sun$, 
which would be valid within a factor of few. 
We then consider three cases:
 (1) $n_{\rm H} = 10\ {\rm cm}^{-3}$ and $f=0.6$; 
 (2) $n_{\rm H} = 100\ {\rm cm}^{-3}$ and $f=0.06$; 
 (3) $n_{\rm H} = 1000\ {\rm cm}^{-3}$ and $f=0.006$. 
Note that the constant product of $fn_{\rm H}$ implies 
the fixed mass in the gamma-ray-emitting region. 
Case (1) is considered for a reference purpose, even though it would hardly explain 
 the similarity between  the synchrotron and the [Fe II] images. 
This set of parameters is more 
appropriate for the X-ray emitting gas, whose density is estimated as 
$n \sim 5\mbox{--}8\ {\rm cm}^{-3}$ \citep{Miceli}.  

%%%%%%%%%%%%%%%%%%%%%%%%%%%%%%%%%%%%%%%%%%

%%%%%%%%%%%%%%%%%%%%%%%%%%%%%%%%%%%%%%%%%%

We adopt the following form as injection distributions of protons and electrons \citep{FermiW51C}:
%%%%%%%%%%%%
% a dummy line  %%%

%%%%%%%%%%%%
\begin{equation}
Q_{e,p}(p)=a_{e,p}\left(\frac{p}{p_0}\right)^{-s}\left(1+\left(\frac{p}{p_{\rm br}}\right)^2\right)^{-\Delta s/2},  
\end{equation}
where $p_0 = 1\ {\rm GeV}\, c^{-1}$. 
The indices and the break momentum are set to be common between 
electrons and protons. 
The radio synchrotron index $\alpha = 0.48$ \citep{radioSNR} corresponds to $s \simeq 2$. 
 The kinetic equation for the momentum distribution of high-energy particles 
 in the shell can be written as:  
 %%%%%%%%%%%%
% a dummy line  %%%

%%%%%%%%%%%%
\begin{equation}
\frac{\partial N_{e,p}}{\partial t}  = \frac{\partial}{\partial p} ( b_{e,p} N_{e,p})  + Q_{e,p},
\end{equation}
where $b_{e,p}=-dp/dt$ is the momentum loss rate, 
 and $Q_{e,p}(p)$ 
(assumed to be time-independent) is the particle injection rate. 
To obtain the radiation spectra from the remnant, 
$N_{e,p}(p, T_0)$ is numerically calculated for $T_0= 2000$~yr. 
Note that energy loss processes such as ionization/Coulomb and synchrotron losses 
are generally not fast enough to modify the gamma-ray spectrum in the LAT band. 
The gamma-ray emission mechanisms include the $\pi^0$-decay gamma rays due to 
high-energy protons, and bremsstrahlung and IC scattering processes by high-energy 
electrons. 
Calculations of the gamma-ray emission were done using the method described 
in \citet{FermiW51C}.
The large gamma-ray luminosity of $L_{\gamma} \sim 1\times 10^{36}\ {\rm erg\ s}^{-1}$ 
precludes IC scattering as a dominant contributor to the gamma-ray emission 
as discussed in \citet{FermiW51C}. Specifically, the total energy required in electrons 
would be unrealistically large $W_e = \int (\gamma -1)m_ec^2  N_e dp \sim 1\times 10^{51}\ {\rm erg}$. 
We shall consider the $\pi^0$-decay and electron bremsstrahlung models 
to account for the observed gamma-ray spectrum.

%%%%%%%%%%%%%%%%%%%%%%%%%%%%%%%%%%%%

%\begin{turnpage}
\begin{deluxetable*}{lcccccccccc}
\tabletypesize{\scriptsize}
%\rotate
\tablecaption{Parameters of Multiwavelength models \label{tbl:model}}
\tablewidth{0pt}
\tablehead{
\colhead{} & \multicolumn{6}{c}{Parameters} & \colhead{}
& \multicolumn{3}{c}{Energetics} \\
\cline{2-7} \cline{9-11} \\
\colhead{Model} & \colhead{$a_e/a_p$} & \colhead{$\Delta s$}
& \colhead{$p_{\rm br}$}  & \colhead{$B$} & \colhead{${n}_{\rm H}$} & \colhead{$f$} 
& \colhead{}
& \colhead{(a)$W_p$ or (b)$W_e$ } 
& \colhead{(a)$U_p$ or (b)$U_e$} & \colhead{$U_B$}\\
\colhead{} & \colhead{} & \colhead{} 
& \colhead{(GeV $c^{-1}$)}  & \colhead{($\mu$G)} & \colhead{(cm$^{-3}$)} & \colhead{} 
& \colhead{} 
& \colhead{($10^{50}$ erg)} & \colhead{(eV cm$^{-3}$)} & \colhead{(eV cm$^{-3}$)} 
}
\startdata
(Case a1) $\pi^0$-decay 
& 0.01 & 0.7 & 4 & 15 & 10 & 0.6 & 
& 11 & $1.1\times 10^5$ & 5.6 \\
(Case a2) $\pi^0$-decay 
& 0.01 & 0.7 & 4 & 60 & 100 & 0.06 & 
& 1.1 & $1.1\times 10^5$ &  90 \\
(Case a3) $\pi^0$-decay 
& 0.01 & 0.7 & 4 & 240 & 1000 & 0.006 & 
& 0.10 & $1.0\times 10^5$ &  1400 \\
(Case b1) Bremsstrahlung 
& 1.0 & 1.0 & 4 & 5 & 10 & 0.6 &
& 2.6 & $2.6\times 10^4$ & 0.62\\
(Case b2) Bremsstrahlung 
& 1.0 & 1.0 & 4 & 20 & 100 & 0.06 & 
& 0.23 & $2.3\times 10^4$ & 10 \\
(Case b3) Bremsstrahlung 
& 1.0 & 1.0 & 4 & 80 & 1000 & 0.006 &
& 0.016 &  $1.6\times 10^4$ & 160 
\enddata
\tablecomments{
Seed photons for IC include infrared 
($kT_{\rm IR} = 3\times 10^{-3}$ eV, $U_{\rm IR} = 1\ \rm eV\ cm^{-3}$), 
optical  ($kT_{\rm opt} = 0.25$ eV, $U_{\rm opt} = 1\ \rm eV\ cm^{-3}$),
and the CMB. The total energy, $W_{e,p}$ and energy density, $U_{e,p}$ of radiating particles are calculated for $p > 10$~MeV~c$^{-1}$.}
\end{deluxetable*}
%\end{turnpage}
%%%%%%%%%%%%%%%%%%%%%%%%%%%%%%%%%%%%

%%%%%%%%%%%%%%%%%%%%%%%%%%%%%%%%%%%%
\begin{figure}[htbp] 
\epsscale{1.0}
\plotone{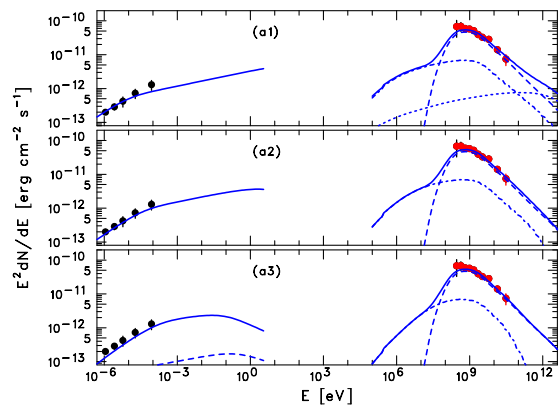}
\caption{\small 
SEDs of W49B with model curves for three cases. Cases a1, a2 and a3 represent $n_{\rm H}=10, 100, 1000\ {\rm cm}^{-3}$, respectively (see Table~\ref{tbl:model}). The gamma-ray emission is assumed to be dominated by $\pi^0$-decay. The radio emission \citep{Moffett} is explained by synchrotron radiation from primary and secondary electrons. The dashed line in the radio band represents the synchrotron emission from the secondary electrons.
The gamma-ray emission is modeled with a combination of $\pi^0$-decay (dashed line), bremsstrahlung (dot-dashed line) and IC scattering (dotted line).
\label{fig:mw_pi0}
}
\end{figure}
%%%%%%%%%%%%%%%%%%%%%%%%%%%%%%%%%%%%

The SED of SNR W49B in the radio and gamma-ray bands is shown in Figure~\ref{fig:mw_pi0}, 
together with the $\pi^0$-decay emission models. 
The radio data are modeled by the synchrotron radiation. 
We construct the $\pi^0$-decay emission models for the different values of 
$n_{\rm H}=10, 100, 1000\ {\rm cm}^{-3}$ (Table~\ref{tbl:model}).
Leptonic components (synchrotron, bremsstrahlung, and IC) are calculated assuming 
$a_e/a_p = 0.01$, a value similar to what is observed for cosmic rays at GeV energies. 
%%%%%%%%%%%%%%%%%%%%%%%%%%%%%
Note that contributions of 
the secondary electrons and positrons produced in {\it pp} collisions are small for the sets of parameters that we adopted~\citep[Table1; see also][]{FermiW44}. 
The secondary synchrotron spectrum is shown in Figure~\ref{fig:mw_pi0} (a3), 
where its flux is about 10\% of the total synchrotron flux at 1 GHz for $n_{\rm H} = 1000\ {\rm cm}^{-3}$. 
The contribution of the secondaries to the gamma-ray emission is also small, about 10\% of the electron bremsstrahlung components for $n_{\rm H} = 1000\ {\rm cm}^{-3}$.

 %%%%%%%%%%%%%%%%%%%%%%%%%%%%%
The product of $n_{\rm H}$ and $W_p$ remains 
almost constant irrespective of $n_{\rm H}$: 
$n_{\rm H}W_p \simeq 10\times 10^{51}$~erg~cm$^{-3}$.
We obtain $B \simeq 240\ \mu$G in the case of  $n_{\rm H} = 1000\ {\rm cm}^{-3}$. 
The SED itself can be formally explained in all the cases. 
The energy density of relativistic protons amounts to $U_p \simeq 1 \times 10^5\ {\rm eV~cm}^{-3}$. 
This value is much higher than $U_p \sim 100\ {\rm eV~cm}^{-3}$ calculated for $\pi^0$-decay dominant modeling of middle-aged SNR W51C~\citep{FermiW51C}.

%%%%%%%%%%%%%%%%%%%%%%%%%%%%%%%%%%%%
\begin{figure}[htbp] 
\epsscale{1.0}
\plotone{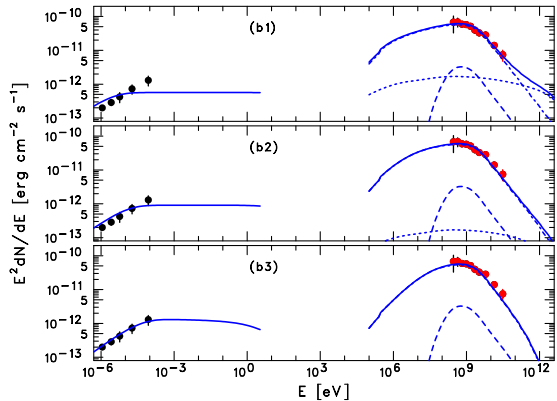}
\caption{\small 
Same as Fig.~\ref{fig:mw_pi0} but the gamma-ray emission is assumed to be dominated by electron bremsstrahlung. 
Cases b1, b2 and b3 represent $n_{\rm H}=10, 100, 1000\ {\rm cm}^{-3}$, respectively (see Table~\ref{tbl:model}).
\label{fig:mw_brems}
}
\end{figure}
%%%%%%%%%%%%%%%%%%%%%%%%%%%%%%%%%%%%

In Figure~\ref{fig:mw_brems}, the gamma-ray spectrum is modeled formally by relativistic bremsstrahlung of electrons. The less-luminous $\pi^0$-decay component is also plotted using $a_e/a_p = 1$. 
It is shown that $n_{\rm H} \ga 100\ {\rm cm}^{-3}$ is required to reproduce the radio spectrum. 
If relativistic bremsstrahlung is responsible for the gamma rays, the ratio of the energy density of relativistic electrons to that of magnetic fields becomes very high, $U_e/U_B  \ga 100$ (Table~\ref{tbl:model}).
The energy density required ($\simeq 2 \times 10^4\ {\rm eV~cm}^{-3}$) is also much higher than $U_e \sim 20\ {\rm eV~cm}^{-3}$ calculated for  bremsstrahlung dominant modeling of W51C.

\section{Conclusions}

We have studied gamma-ray emission in the direction of SNR W49B using about 17 months of data accumulated by the \emph{Fermi} LAT. 
The observed energy spectrum in 0.2--200~GeV exhibits steepening toward high energies, although a simple power-law function cannot be completely ruled out given the uncertainties of the Galactic diffuse model. 
The luminosity is estimated to be $1.5 \times 10^{36}~(D/8$~kpc)$^2$~erg~s$^{-1}$, which makes this source one of 
the most luminous gamma-ray sources in the Galaxy. 

The gamma-ray source is unresolved by the LAT, which is consistent with the angular size 
of SNR W49B ($\sim 4\arcmin$ in diameter) taking into account the effective LAT PSF. Assuming a point source, the source position is found to be ($\alpha$, $\delta$)=($287\fdg 756$, $9\fdg 096$) with an error radius of $0\fdg 024$ at 95\% confidence level. This result clearly shows that the gamma-ray emission comes predominantly from the SNR W49B region, not from a nearby star forming region, W49A. We put an upper limit on the gamma-ray luminosity of W49A as 
$< 3\times 10^{35}\ (D/11.4\ {\rm kpc})^2\ {\rm erg\ s}^{-1}$ at 95\% confidence level.
The gamma-ray emission in the direction of SNR W49B 
is unlikely to come from a pulsar. The gamma-ray energy distribution is different from that observed for other pulsars with the LAT. 
In addition, no pulsed emission has been detected with the LAT nor are any radio pulsars known in this direction.

A good match between the best-fit LAT position and the brightest part of synchrotron radio emission suggests that the gamma-ray source originates in the shell of SNR W49B. 
The LAT spectrum can be formally explained either by $\pi^0$-decay gamma rays or by electron bremsstrahlung.
For both cases, the calculated energy density of relativistic particles is evaluated to be very high, $U_{e,p}>10^4\ {\rm eV\ cm}^{-3}$.

\acknowledgments
The \textit{Fermi} LAT Collaboration acknowledges generous ongoing support
from a number of agencies and institutes that have supported both the
development and the operation of the LAT as well as scientific data analysis.
These include the National Aeronautics and Space Administration and the
Department of Energy in the United States, the Commissariat \`a l'Energie Atomique
and the Centre National de la Recherche Scientifique / Institut National de Physique
Nucl\'eaire et de Physique des Particules in France, the Agenzia Spaziale Italiana
and the Istituto Nazionale di Fisica Nucleare in Italy, the Ministry of Education,
Culture, Sports, Science and Technology (MEXT), High Energy Accelerator Research
Organization (KEK) and Japan Aerospace Exploration Agency (JAXA) in Japan, and
the K.~A.~Wallenberg Foundation, the Swedish Research Council and the
Swedish National Space Board in Sweden.

Additional support for science analysis during the operations phase is gratefully
acknowledged from the Istituto Nazionale di Astrofisica in Italy and the Centre National d'\'Etudes Spatiales in France.

%% The following command ends your manuscript. LaTeX will ignore any text
%% that appears after it.

\end{document}